\def\aa{{A\&A}}

\def\aj{{AJ}}
\def\annrev{{ARA\&A}}
\def\apj{{ApJ}}

\def\mnras{{MNRAS}}
\def\nat{{Nature}}

\documentclass[cup5b]{caps}
\usepackage{graphicx}
\usepackage{amssymb}
\usepackage{psfig}
\usepackage{ociwsymp1}
\HeadText{M. C. Begelman}

\def\plotone#1{\centering \leavevmode
\includegraphics[width=.95\columnwidth]{#1}}

\begin{document}

\pagenumbering{arabic}

\author[]{M. C. BEGELMAN\\JILA, University of Colorado at Boulder}

\chapter{AGN Feedback Mechanisms}

\begin{abstract}
Accreting black holes can release enormous amounts of energy to their 
surroundings, in various forms. Such feedback may profoundly influence a black 
hole's environment. After briefly reviewing the possible types of feedback, I 
focus on the injection of kinetic energy through jets and powerful winds.  The 
effects of these outflows may be especially apparent in the heating of the 
X-ray--emitting atmospheres that pervade clusters of galaxies. Analogous 
heating effects, during the epoch of galaxy formation, could regulate the 
growth of supermassive black holes.
\end{abstract}

\section{Introduction}

Active galactic nuclei (AGNs) release large amounts of energy to their 
environments, in several forms.  In luminous AGNs such as Seyfert nuclei and 
quasars, the most obvious output is radiative; indeed, radiation can affect 
the environment through both radiation pressure and radiative heating.  
Although jets and winds are usually associated with radio galaxies, 
recent theoretical and observational developments suggest that the kinetic 
energy output may be as important as (or more important than) the 
radiative output for most accreting black holes. Finally, significant 
outputs of energetic particles, whether charged (``cosmic rays") or 
neutral (relativistic neutrons, neutrinos), cannot be ruled out.

In this review, I focus on the effects of kinetic energy feedback, since these 
are probably most relevant to the coevolution of black holes and galaxies.  
Recent observations of galaxy clusters suggest that AGN feedback plays a 
crucial role in regulating the thermodynamics of the intracluster medium 
(ICM).  I discuss the nature of this interaction, then extrapolate to similar 
effects that may have operated in protogalaxies, during the era when 
supermassive black holes were growing toward their present masses. 

\section{Forms of Feedback}

Before specializing to the case of kinetic energy injection in cluster 
atmospheres, I briefly review the various forms of energy injection and 
summarize their likely effects.  This section is an updated version of 
the discussion given in Begelman (1993).

\subsection{Radiation Pressure}

Radiation pressure can exert a force on the gas via electron scattering, 
scattering and absorption on dust, photoionization, or scattering in 
atomic resonance lines.  Electron scattering is the simplest mechanism to 
treat, with a cross section  of 
$\langle \sigma / H \rangle \sim 7\times 10^{-25} x$ cm$^2$ per hydrogen atom, 
where $x$ is the ionized fraction.  The maximum column density over which the 
force can be exerted is given by $N_{H,max} \sim  
\langle \sigma / H \rangle^{-1} \sim 2\times 10^{24} x^{-1}$ cm$^{-2}$.

If the radiation flux from the nucleus does not greatly exceed the Eddington 
limit of the central black hole, the radiation force exerted through electron 
scattering will have a relatively minor dynamical effect on the gas in 
the host galaxy, compared to gravitational and thermal pressure forces.  In 
contrast, radiation pressure acting on dust can exert a much larger force 
per H atom, although over a correspondingly smaller column density.  If we 
assume a dust-to-gas ratio (by mass) of 0.01, a typical grain size $a$, and a 
total cross section per grain of the same order as the geometric cross section 
(a reasonable assumption for UV and soft X-ray photons hitting 
$\sim 0.1 \ \mu$m grains), then the cross section per H atom exceeds that 
for electron scattering in fully ionized gas by a factor of order 
$10^5 (a/ 0.1 \ \mu{\rm m})^{-1}$.  The force exerted per particle is higher 
by the same factor, but the column density affected is only $2\times 10^{19}  
(a/ 0.1 \ \mu{\rm m})$ cm$^{-2}$. Dopita and collaborators have argued that 
this form of pressure could dominate the dynamics of narrow emission-line 
regions in AGNs, under certain conditions, and could be crucial for regulating 
the ionization state (Dopita et al. 2002; Dopita 2003).

The force exerted on $\sim 10^4 - 10^5$ K gas as a result of steady-state 
photoionization and recombination is characterized by the mean cross section 
$\langle \sigma / H \rangle \sim 10^{-18} (1-x)$ cm$^2$.  Photoionization 
equilibrium predicts that $(1-x) \sim 10^{-4} p_{gas}/p_{rad}$, where 
$p_{gas}$ and $p_{rad}$ are the gas pressure and pressure of ionizing
radiation, respectively. (Strictly speaking, one should use $4\pi J/c$ instead 
of $p_{rad}$, where $J$ is the mean intensity.  However, use of $p_{rad}$ is 
quantitatively correct and promotes a more physically intuitive discussion.)  
Thus, we can write $\langle \sigma / H \rangle \sim 
10^{-22} p_{gas}/ p_{rad}$ cm$^2$.  Note that $p_{gas}/ p_{rad}$ is just the 
reciprocal of the ``ionization parameter" $\Xi$ defined by Krolik, McKee, \& 
Tarter (1981).  The force per H atom exerted through photoionization is 
$200 p_{gas}/ p_{rad}$ times greater than that exerted through electron 
scattering.  

The largest forces per H atom are possible through scattering in UV resonance 
lines.  For species $i$, the effective cross section is 
\begin{equation}
\left\langle {\sigma \over H } \right\rangle \sim 
{1\over  \Delta\nu_D}{\pi e^2\over  m_ec} (A_iX_if_i),
\end{equation}
where $\Delta\nu_D$ is the Doppler width of the line and $A_i$, $X_i$, and 
$f_i$ are, respectively, the abundance of element $i$, the fraction of the 
element in the relevant ionization state, and the oscillator strength of the 
transition.  For important resonance lines such as those of C~IV, Si~IV, and 
N~V, $A_iX_if_i$ can attain values of order $10^{-4}$ for cosmic abundances.  
The fractional Doppler width, at $T\sim 10^4$ K, is 
$\Delta\nu_D/ \nu \sim 10^{-4}$; hence, we find that the mean cross section 
can be as large as $10^{-17}$ cm$^2$ and the corresponding force as much as 
{\it seven orders of magnitude} larger than electron scattering.  The drawback 
is that the bandwidth over which resonance-line scattering is effective is 
extremely small, fractionally of order $\sim 10^{-4}$ of the ionizing spectrum 
for each strong resonance line.  The amount of momentum available from such 
a small bandwidth is very small.  Therefore, in order for resonance-line 
scattering to be dynamically important, (1) the gas must accelerate, so that 
new portions of the spectrum are continuously Doppler shifted into the line 
(the basis for the Sobolev approximation), and (2) there must be a significant 
number of lines contributing at different wavelengths.  Models of UV 
resonance-line acceleration for O-star winds (Castor, Abbott, \& Klein 1975a) 
have been adapted to explain the fast ($\upsilon \rightarrow 0.1 c$) outflows 
in broad absorption-line (BAL) QSOs (Arav \& Li 1994; Arav, Li, \& Begelman 
1994; Murray et al. 1995), where there is circumstantial evidence for 
acceleration by radiation pressure (Arav 1996; Arav et al. 1999). 
 
For AGN radiation acting on general interstellar matter, the radiation 
pressure force will be exerted mainly through dust and photoionization.  
Dynamical effects can be significant if $p_{rad} > p_{ISM}$, where $p_{ISM}$ 
is the pressure in the undisturbed gas and 
\begin{equation} 
p_{rad} = {L\over 4\pi R^2 c} = 3\times 10^{-9} L_{46} R_{kpc}^{-2} \ 
{\rm dyne \ cm}^{-2}
\end{equation} 
for gas situated $R_{kpc}$ kpc from an isotropic source of ionizing radiation 
with luminosity $10^{46} L_{46}$ erg s$^{-1}$.   

The maximum column density in a slab of gas that can be fully ionized by AGN 
continuum is given by
\begin{equation}
N_{H,ion} \sim 10^{22} \ln \left(1+ {p_{rad}\over p_{ISM}}\right) \ 
{\rm cm}^{-2}  
\end{equation}  
if dust absorption is neglected, and only slightly smaller if a cosmic dust 
abundance is taken into account.  If $p_{ISM} < p_{rad}$, then $N_{H, ion}$ 
gives the depth to which an irradiated cloud will be ``pressurized" by the 
radiation.  Differential pressure forces between the front and back of a cloud 
can lead to a ``pancake effect" (Mathews 1982), which tends to squash 
irradiated clouds down to a column density of order $N_{H,ion}$.   

\subsection{Radiative Heating}

Gas exposed to ionizing radiation from an AGN tends to undergo an abrupt 
transition from the typical H~II region temperature, $\sim 10^4$ K, to a 
higher temperature and ionization state when $p_{gas}/p_{rad}$ falls below 
some critical value, which lies in the range $\sim 0.03-0.1$ (McCray 1979; 
Krolik et al. 1981).  The ``hot phase" equilibrium temperature is close 
to the temperature at which Compton cooling balances inverse Compton heating, 
which  can be $T_{IC} \sim  10^6 - 10^7$ K for AGN spectra.  
 
Clouds with column densities greater than $N_{H,ion}$ will not heat up all at 
once.  Only the surface layers will be ablated, and the back-pressure of the 
ablated gas will keep the cloud interior at a high enough pressure to avoid 
immediate heating (Begelman, McKee, \& Shields 1983; Begelman 1985).  
Depending on the size of the cloud and the timescales involved, the heated gas 
may reach a temperature $T_s\sim 10^5- 10^6$ K at the point at which it 
becomes supersonic with respect to the cloud surface.  A critical condition 
at the sonic point determines the mass flux per unit area, which is 
proportional to $p_{rad}/ T_s^{1/2}$:    
\begin{equation}
\dot N \sim (6\times 10^6 - 6\times 10^7) {L_{46}\over R^2_{kpc}} \ 
{\rm cm^{-2} \ s^{-1}}.
\end {equation}
The lifetime of a cloud against ablation by X-ray heating is given by
\begin{equation}
{N\over \dot N} \sim (5\times 10^6 - 5\times 10^7) { R^2_{kpc} \over L_{46}} 
{N_H\over 10^{22} \ {\rm cm^{-2}}}  \ {\rm yr} ,
\end{equation}
which corresponds to a global ablation rate of
\begin{equation}
\dot M_{abl} \sim (20-200) {\cal C} L_{46} \ M_\odot \ {\rm yr}^{-1}
\end{equation}
for gas with a covering factor ${\cal C}$.  The effects of X-ray heating on 
the evolution of the ISM in a spiral galaxy were studied by Begelman (1985) 
and by Shanbhag \& Kembhavi (1988).  Possible observable consequences of X-ray 
heating include: (1) elimination of cool ISM phases from the inner parts of 
the galaxy, which would allow UV radiation to penetrate to much larger 
distances than would otherwise be possible; (2) modification of ISM phase 
structure by preferential destruction of small clouds; only clouds (e.g., 
giant molecular clouds) with a sufficiently large column density would survive 
long enough to be observed; and (3) generation of peculiar cloud velocities  
$> 100$ km s$^{-1}$, via the ``rocket effect," with a random component
introduced through cloud-cloud shadowing.   The importance of these effects 
is highly uncertain as they depend critically on the geometric distribution 
of the dense phases of the ISM, and on the replenishment of ISM through 
stellar evolutionary and other processes.

\subsection{Energetic Particles}

In addition to charged relativistic particles that can diffuse through the 
surrounding medium (``cosmic rays"), AGNs may also emit ``exotic" particle 
outflows, consisting, for example, of relativistic neutrons and neutrinos 
(Begelman, Rudak, \& Sikora 1990).  These would tend to deposit their energy 
in a volume-distributed fashion, rather than in impulsive fashion at a shock 
front. Even neutrinos, if they are sufficiently energetic --- TeV or above --- 
could be absorbed by nearby stars and heat their interiors (Czerny, Sikora, \& 
Begelman 1991).

Ultrarelativistic neutrons could have particularly interesting effects since 
relativistic time dilation would allow them to travel large distances 
unimpeded before they decay and couple to the ambient plasma (Sikora, 
Begelman, \& Rudak 1989).  This form of energy injection could drive powerful, 
fast winds that start far from the central engine (e.g., Begelman, de Kool, 
\& Sikora 1991).  We note, however, that to date there is no compelling 
evidence that most AGNs emit a large fraction of their energy in this form. 
Dynamical effects of neutron winds have been invoked recently to explain 
certain features of gamma-ray burst afterglows (Beloborodov 2003; Bulik, 
Sikora, \& Moderski 2003).

\subsection{Kinetic Energy}

In addition to the obvious example of radio galaxies, which often release the bulk of their power in the form of jets (Rees et al. 1982; Begelman, Blandford, \& Rees 1984), most if not all accreting black holes could produce substantial outflows.  Numerical simulations suggest that accretion disks, which transfer angular momentum and dissipate binding energy via magnetorotational instability, may inevitably produce magnetically active coronae (Miller \& Stone 2000).  These likely generate outflows that are further boosted by centrifugal force (Blandford \& Payne 1982). We have already mentioned the possible role of radiation pressure in accelerating winds in BAL QSOs --- such outflows could also be boosted 
hydromagnetically.  Whatever the acceleration mechanism(s), these winds 
are probably accelerated close to the central engine, and therefore should be 
regarded as part of the kinetic energy output.  New spectral analyses of 
BAL QSOs, made possible by observations with the {\it Hubble Space Telescope}, 
imply that the absorption can be highly saturated (Arav et al. 2001) and may 
originate far from the nucleus (de Kool et al. 2001). This indicates that the 
kinetic energy in the BAL outflow is larger than previously thought and can 
approach the radiation output.  Moreover, new evidence suggests that 
relativistic jets are common or ubiquitous in X-ray binaries containing black 
hole candidates.  While the energetics of these outflows are not yet fully 
established, their environmental impacts may be substantial (Heinz 2002). 

Unless radiation removes at least 2/3 of the liberated binding energy, very 
general theoretical arguments indicate that rotating accretion flows 
{\it must}\ lose mass.  The physical reason is that viscous stresses transport 
energy outward, in addition to angular momentum.  If radiation does not remove 
most of this energy, then a substantial portion of the gas in the flow will 
gain enough energy to become unbound (Narayan \& Yi 1995; Blandford \& 
Begelman 1999). Blandford (this volume) discusses this effect and its 
consequences in more detail. While it may sometimes be possible to tune the 
system so that the gas circulates without escaping, any excess dissipation 
(i.e., increase of entropy) near a free surface of the flow will lead to 
outflow.  There are several possible sources of such dissipation, including 
magnetic reconnection, shocks, radiative transport, and the magnetocentrifugal 
coupling mentioned above. If radiative losses are very inefficient, outflows 
can remove all but a small fraction of the matter supplied at large radii.   

\section{Energy Budget}

At an energy conversion efficiency of $\epsilon c^2$ per unit of accreted 
mass, an accreting black hole liberates $10^{19} (\epsilon/0.01)$ erg per 
gram. In principle the efficiency could be $\sim 6$ to more than $40$ times 
larger than this (depending on the black hole spin and boundary conditions 
near the event horizon: Krolik 1999; Agol \& Krolik 2000), but we have chosen 
deliberately to be conservative.  $\epsilon$ might be viewed as the efficiency 
of kinetic energy production, since this is probably the most effective means 
by which black holes affect their surroundings.  In a galactic bulge with a 
velocity dispersion of $200 \sigma_{200}$ km s$^{-1}$, the accretion of one 
gram liberates enough energy to accelerate $2 \times 10^4 (\epsilon/0.01) 
\sigma_{200}^2 $ gm to escape speed --- provided that most of the energy goes 
into acceleration.  Given a typical ratio of black hole mass to galactic bulge 
mass of $\sim 10^{-3}$, feedback from a supermassive black hole growing toward 
its final mass could easily exceed the binding energy of its host galaxy's 
bulge.   

Under many circumstances, feedback via kinetic energy injection can be quite 
efficient.  Both radiative heating and acceleration by radiation pressure, on 
the other hand, have built-in inefficiencies. In both photoionization heating 
and Compton heating, only a fraction of the photon energy goes into heat, the 
majority being reradiated. For example, the Compton heating efficiency per 
scattering is $\sim kT_{IC} / m_e c^2 < 10^{-2}$ for typical AGN Compton 
temperatures.  Acceleration by radiation pressure extracts only a fraction 
$\sim \upsilon/c$ of the available energy per scattering, where $\upsilon$ is 
the speed of the accelerated gas. For outflows in BAL QSOs, with velocities of 
up to $\sim 0.2 c$, this can represent an efficient energy injection 
mechanism, even for single scattering.  The energetic efficiency of radiative 
acceleration is increased if the photons scatter multiple times, to 
$\tau \upsilon/c$, where $\tau$ is the optical depth.  

Even the efficacy of kinetic energy injection depends on the structure of the 
medium in which it is deposited. The speed of a shock or sound wave 
propagating through a medium with a ``cloudy" phase structure will be highest 
in the phase with the lowest density (the intercloud medium).  Dense regions 
will be overrun and left behind by the front, as first pointed out by McKee \& 
Ostriker (1977) in connection with supernova blast waves propagating into the 
interstellar medium. Consequently, most of the energy goes into the gas 
which has the lowest density (and is the hottest) to begin with.  The global 
geometric structure of the ambient gas is important as well.   Since a wind 
or hot bubble emanating from an AGN will tend to follow the ``path of least 
resistance," a disklike structure can lead to a ``blowout" of the hot gas 
along the axis.  A more spherically symmetric gas distribution will tend to 
keep the AGN energy confined in a bubble. 

\section{AGN Feedback in Clusters}

AGN feedback due to kinetic energy injection is perhaps most evident in 
clusters of galaxies.   Recent observations of intracluster gas by 
{\it Chandra}\ and {\it XMM-Newton}\ indicate that some energy source is 
quenching so-called ``cooling flows" in clusters of galaxies (Allen et al. 
2001; Fabian et al. 2001; Peterson et al. 2001).  Energy injected by 
intermittent radio galaxy activity at the cluster center is the most likely 
culprit.  The same form of energy input, spread over larger scales, could be 
responsible for an inferred ``entropy floor" in the gas bound to clusters of 
galaxies (Valageas \& Silk 1999; Nath \& Roychowdury 2002). 

\subsection{Evolution of Radio Galaxies}

Radio galaxies evolve through three stages, only the first of which 
is dominated by the jet momentum. Although the radio morphologies in powerful 
(Fanaroff-Riley class II) sources are dominated by the elongated lobes and 
compact hotspots, most of the energy accumulates in a faint ``cocoon" that has 
a thick cigar shape (Blandford \& Rees 1974; Scheuer 1974).  The same is 
probably true of the weaker FR~I sources, which appear to be dominated by 
emission from turbulent regions along the jet. The cocoon is overpressured 
with respect to the ambient medium, and drives a shock ``sideways" at the 
same time as the jets are lengthening their channels by depositing momentum. 
The sideways expansion quickly becomes competitive with the lengthening. The 
archetypal FR~II source Cygnus A, for example, which appears to be long and 
narrow in the radio, displays an aspect ratio $<3$ in X-rays (although the 
hotspots remain prominent in the X-ray image: Wilson, Young, \& Shopbell 2000).

Dynamically, active radio galaxies with overpressured cocoons resemble 
spherical, supersonic stellar wind bubbles (Castor, McCray, \& Weaver 1975b; 
Begelman \& Cioffi 1989). To zeroth order, the evolution of the bubble 
can be described by a self-similar model in which the internal and kinetic 
energy are comparable, and share the integrated energy output of the wind.  
The speed of expansion is then
\begin{equation}
\upsilon \sim \left({L_j\over \rho}\right)^{1/5} R^{-2/3},
\end{equation}
where $L_j$ is the power of the jets, $\rho$ is the ambient density, and $R$ 
is the radius of the shock.  The supersonic expansion phase ends when the 
expansion speed drops below the sound speed in the ambient medium. This occurs 
at a radius
\begin{equation}
R_{\rm sonic}\sim 5 \left({\langle L_{43} \rangle \over n}\right)^{1/2} 
T_{\rm keV}^{-1/4} \ {\rm kpc} , 
\end{equation}
where $\langle L_{43} \rangle$ is the time-averaged jet power in units of 
$10^{43}$ erg s$^{-1}$, $n$ is the ambient particle density in units of 
cm$^{-3}$, and $T_{\rm keV}$ is the ambient temperature in units of keV.  
Thereafter the evolution is dominated by buoyancy (Gull \& Northover 1973).  
We have chosen fiducial parameters that are fairly typical of conditions in 
cD galaxies at the centers of rich clusters --- note how small $R_{\rm sonic}$ 
is, compared to a typical cluster core radius, or even the core radius of the 
host galaxy.  Cygnus A, which has been expanding for several million years, is 
hundreds of kpc across, and is still overpressured by a factor $\sim 2-3$ with 
respect to the ambient medium, is the exception rather than the rule.  It is 
a very powerful source expanding into a relatively tenuous ambient medium 
(Smith et al. 2002). In the X-ray emitting clusters that we discuss below the 
central radio galaxies seem to have evolved into the buoyancy-driven stage 
fairly early.

At least two additional caveats must be taken into account in considering the 
effects of radio galaxies on their surroundings.  First, the active production 
of jets is probably intermittent --- there is indirect statistical evidence 
that the duty cycle may be as short as $10^5$ yr (Reynolds \& Begelman 1997).  
During ``off" periods, the overpressured bubble continues to expand as a 
blast wave with fixed total energy, but the radio emissivity may rapidly 
fade.  Second, the direct influence of the radio galaxy on its surroundings 
does not end with the onset of buoyancy-driven evolution. As I describe below, 
the buoyant bubbles of very hot (possibly relativistic) plasma seem fairly 
immiscible with their surroundings.  They can ``rise" for considerable 
distances, spreading the AGN's energy output widely.  Both of these points 
will figure prominently in the next section.

\subsection{Quenching of Cooling Flows}

Radiative cooling of gas in the central regions of galaxy clusters often 
occurs on a timescale much shorter than the Hubble time.  In the absence of 
any heat sources, this implies that the ICM must settle subsonically 
toward the center in order to maintain hydrostatic equilibrium with the gas at 
larger radii. The mass deposition rates predicted by this ``cooling flow" 
model are very high and range typically from 10 to 1000 solar masses per year. 
X-ray observations made prior to the launch of {\it Chandra}\ seemed to be 
broadly consistent with this picture (Fabian 1994). However, the picture has 
{\it not}\ held up well in the era of {\it Chandra}\ and {\it XMM-Newton}. 
Although both gas temperatures and cooling times are observed to decline 
toward cluster cores, new observations show a remarkable lack of emission 
lines from gas at temperatures below $\sim 1$ keV in the central regions of 
clusters (Allen et al. 2001; Fabian et al. 2001; Peterson et al. 2001), 
suggesting a temperature ``floor" (Peterson et al. 2003) at about $1/2-1/3$ 
of the temperature at the ``cooling radius" (where the  cooling timescale 
equals the Hubble time). Moreover, the mass deposition rates obtained with 
{\it Chandra}\ and {\it XMM-Newton}\ using spectroscopic methods are many 
times smaller than earlier estimates based on {\it ROSAT}\ and {\it Einstein}\ 
observations, as well as more recent morphological estimates based on cooling 
rates alone (David et al. 2001; McNamara et al. 2001; Peterson et al. 2001, 
2003).  The strong discrepancies between these results indicate that the gas 
is prevented from cooling by some heating process. 

The most plausible candidates for heating the ICM are thermal conduction from 
the outer parts of the cluster (Bertschinger \& Meiksin 1986; Narayan \& 
Medvedev 2001; Voigt et al. 2002; Zakamska \& Narayan 2003) and kinetic energy 
injected by a central AGN. Given the metallicities and colors of the galaxies 
hosting cooling flows, heating by supernovae and hot stellar winds seems 
marginally adequate at best (Wu, Fabian, \& Nulsen 2000). AGN heating is 
especially attractive because $\sim 70\%$ of cD galaxies in the centers of 
cooling flow clusters are radio galaxies (Burns 1990). The 
{\it ensemble-averaged}\ power from radio galaxies is more than sufficient to 
offset the mean level of cooling (Peres et al. 1998; B\"ohringer et al. 2002), 
although not every cluster shows strong radio galaxy activity at the present 
time. Moreover, AGN heating is naturally concentrated toward the center of the 
cluster, where the risk of runaway cooling (``cooling catastrophe") 
is greatest. 

It is one thing to argue, on energetic grounds, that AGN feedback is capable 
of replenishing the heat lost to radiation in cooling flows.  It is quite 
another to determine how this happens in detail.  Because of the steep 
dependence of the radiative cooling function on density, it has proven 
notoriously difficult to ``stabilize" cooling flows, so that heating 
approximately balances cooling at all radii. For example, Meiksin (1988) found 
that conduction could not stop cooling catastrophes in the central regions 
of clusters, although it could offset cooling in the outer parts if 
the temperature gradient were not too large.  But even this requires fine 
tuning of the conductivity (via a ``magnetic suppression factor" relative to 
the Spitzer value) and boundary conditions, as too large a conduction rate 
will lead to a nearly isothermal temperature distribution, contrary to 
observations.  Indeed, as Loeb (2002) points out, a large enough conductivity 
to suppress cooling flows, if extrapolated to cluster envelopes, would cause 
them to evaporate.  Early attempts to offset cooling using a central heat 
source (e.g., Loewenstein, Zweibel, \& Begelman 1991, using cosmic rays) ran 
into similar fine-tuning problems. 

The episodic model for cluster heating (Binney \& Tabor 1995; Ciotti \& 
Ostriker 1997, 2001) attempts to avoid these generic difficulties. No steady 
state is sought. Instead, the cluster atmosphere goes through repeated cycles 
of cooling and infall --- which fuel the central AGN --- followed by heating 
and outflow.  The rapid heating and expansion of the ICM turns off the fuel 
supply to the AGN, the initial conditions of the cluster atmosphere are 
``reset," and the process repeats.  The energy injection process, whether due 
to jets (as in Binney \& Tabor) or inverse Compton heating (as in Ciotti \& 
Ostriker), is violent and heats the ICM from the inside out.  This creates an 
observational challenge for this class of models, since they generally predict 
that the temperature should decrease outward during the heating phase --- the 
opposite to what is seen.  Nor are the expected strong shocks observed. If one 
concludes from this that the heating episodes somehow elude observation, the 
same can be said for the cooling catastrophes --- the inevitable conclusion of 
each cooling phase.  These are also not seen, although Kaiser \& Binney (2003) 
point out that they may be sufficiently short-lived to have evaded detection 
in existing datasets. (We also note that the Compton temperatures assumed by 
Ciotti \& Ostriker are based on extreme --- and highly beamed --- spectra of 
blazars, and are probably far too high to be realistic. Using more realistic 
AGN Compton temperatures will weaken this mechanism to the point where it is 
probably not effective. Thus, if episodic mechanisms work at all, it is 
probably through kinetic energy injection.) 

\begin{figure}
\plotone{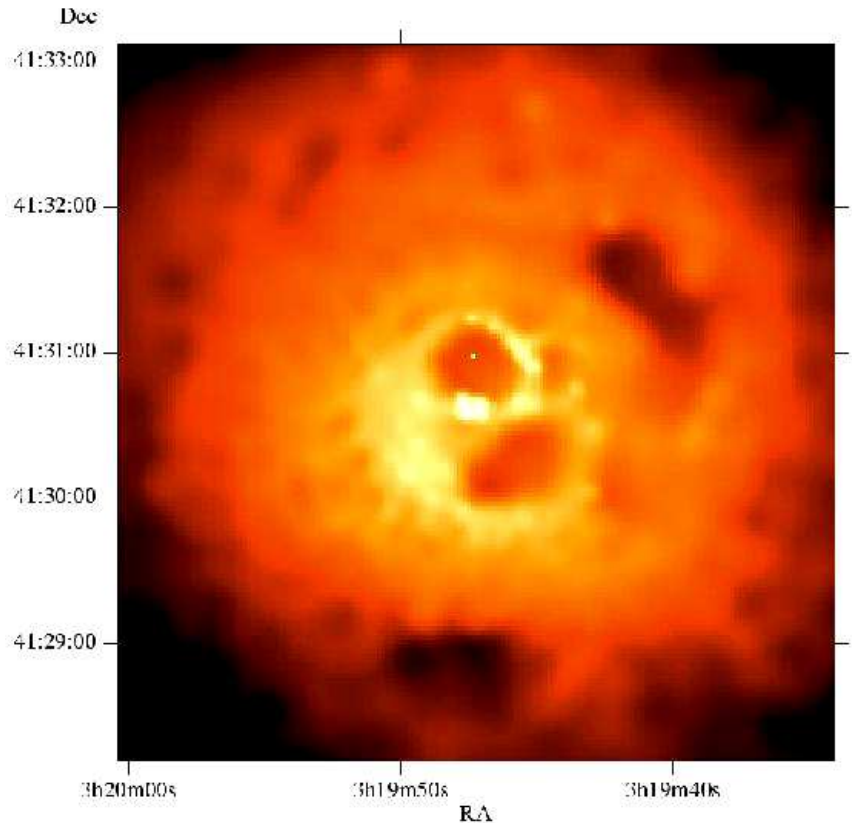}
\caption{Adaptively smoothed, 0.5--7 keV {\it Chandra}\ image of the core 
of the Perseus cluster (from Fabian et al. 2000). The radio lobes of 3C 84 
coincide with the central X-ray holes, which are bounded by bright (cool) 
rims. The more distant ``ghost cavities" are thought to be buoyant bubbles 
created by earlier episodes of activity.}
\label{Fig1}
\end{figure}

The absence of strong shocks bounding radio galaxy lobes is a major 
observational surprise. For example, the X-ray--bright rims surrounding the 
radio lobes of 3C 84 (NGC 1275) in the Perseus cluster are cooler than 
their surroundings (Fabian et al. 2000; Fig.~\ref{Fig1}), contrary to 
predictions (Heinz et al. 1998).  This probably results from the entrainment 
and lifting of low-entropy gas, from the cluster center, into regions where 
the ambient entropy is higher (Reynolds, Heinz, \& Begelman 2001, 2002; 
Quilis, Bower, \& Balogh 2001; Brighenti \& Mathews 2002; Nulsen et al. 
2002).  It is probably not due to the {\it in situ}\ radiative cooling of 
shock-compressed gas bounding the radio lobes.  This shows that 3C 84 has 
already evolved to the buoyancy-driven stage; similar conclusions can be 
drawn for other cluster cores with prominent radio sources.

Another surprise is the apparent ``immiscibility" of the hot (possibly 
relativistic) plasma injected by the jets and the thermal ICM. It has been 
known since the time of {\it ROSAT}\ (B\"ohringer et al. 1993; McNamara, 
O'Connell, \& Sarazin 1996) that the plasma in radio lobes can displace cooler 
thermal gas, creating ``holes" in the X-ray emission.  More sensitive 
{\it Chandra}\ imaging has shown not only how common such holes are, but also 
how long they can persist. In particular, numerous examples of 
``ghost cavities" have been found (e.g., McNamara et al. 2001; Johnstone 
et al. 2002; Mazzotta et al. 2002).  These are presumably buoyant bubbles left 
over from earlier epochs of activity; several examples are seen in 
Fig.~\ref{Fig1}.

The persistence of highly buoyant bubbles may be key to understanding how AGNs 
heat cluster atmospheres.  Strongly positive entropy gradients observed in 
cluster atmospheres (David et al. 2001; B\"ohringer et al. 2002) appear 
to rule out standard convection.  But the Schwarzschild criterion refers to 
heat transport by marginally buoyant fluid elements, not the highly buoyant 
bubbles that appear to be present.  Numerical simulations are beginning to 
address how buoyant plumes of plasma injected by jets can increase the 
potential and thermal energy of the ICM (Quilis et al. 2001; Reynolds et al. 
2001, 2002; Churazov et al. 2002; Br\"uggen \& Kaiser 2002); spread out 
laterally (into ``mushroom clouds": Churazov et al. 2001), yielding more even 
distribution of the injected energy; and persist long after the observable 
radio lobes have faded (Br\"uggen et al. 2002; Reynolds et al. 2002; Basson \& 
Alexander 2003).  The latter point is especially important given statistical 
(Reynolds \& Begelman 1997) and morphological (e.g., Virgo cluster: Young, 
Wilson, \& Mundell 2002; Forman et al. 2003; Perseus cluster: Fabian et al. 
2000, 2002) evidence that typical radio galaxy activity is intermittent, 
possibly with a short duty cycle. Moreover, both radio and X-ray observations 
suggest that the energy ultimately gets distributed remarkably evenly (e.g., 
Owen, Eilek, \& Kassim 2000), despite the apparent immiscibility noted above.  
Whether this mixing is due to the propagation of (magneto-)acoustic waves, 
buoyancy, Kelvin-Helmholtz instabilities, unsteadiness in the jets, mixing by 
``cluster weather" (due, e.g., to galaxy motion or cluster mergers), or other 
effects remains unclear. 

Bubbles rising subsonically do $pdV$ work on their surroundings as they 
traverse the pressure gradient. Since the timescale for the bubbles to cross 
the cluster (of order the free-fall time) is much shorter than the cooling 
timescale, the flux of bubble energy through the ICM approaches a steady 
state, implying that details of the energy injection process --- such as the 
number flux of bubbles (e.g., one big one or many small ones), the bubble 
size, filling factor, and rate of rise --- do not affect the mean heating 
rate. If we assume that the acoustic energy generated by the $pdV$ work is 
dissipated within a pressure scale height of where it is generated, we can 
devise an average volume heating rate for the ICM, as a function of radius 
(Begelman 2001b):  
\begin{equation}
\label{effer}
{\cal H} \sim {\langle L \rangle \over 
4\pi r^3}\left({p\over p_0}\right)^{1/4}\left|{d\ln p \over d\ln r}\right| .
\end{equation}
In eq.~(\ref{effer}), $\langle L \rangle$ is the time-averaged power output of 
the AGN, $p(r)$ is the pressure inside the bubbles (and $p_0$ is the 
pressure where the bubbles are formed), and the exponent $1/4$ equals 
$(\gamma - 1)/\gamma$ for a relativistic plasma (the exponent would be 2/5 
for a nonrelativistic gas).  A major assumption of the model, that the $pdV$ 
work is absorbed and converted to heat within a pressure scale height, will 
have to be assessed using 2D and 3D numerical simulations and studies of the 
microphysics of cluster gas. We have also assumed that the energy is spread 
evenly over $4\pi$ sr, a likely consequence of buoyancy.  

\begin{figure}
\plotone{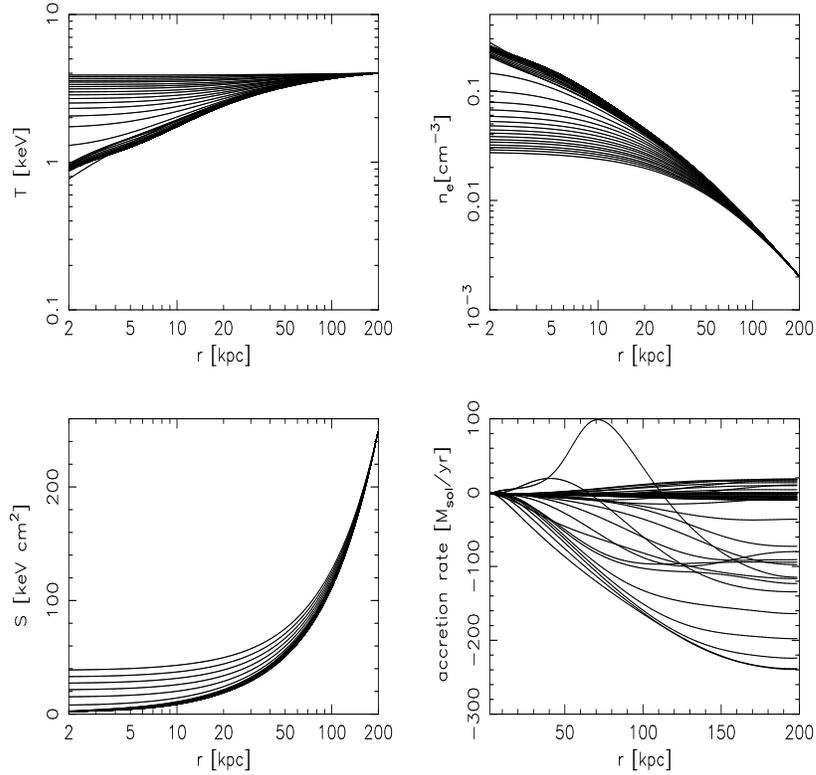}
\caption{ZEUS 1D simulation of cluster evolution over one Hubble time, 
including effervescent heating, thermal conduction, and AGN feedback. The 
figure shows time sequences of (clockwise from upper left) temperature, 
density, accretion rate, and specific entropy profiles.  The model settles 
down to a stable, steady state, which is visible via the dense concentration 
of curves.  See Ruszkowski \& Begelman (2002) for further details.}
\label{Fig2}
\end{figure}

\begin{figure}
\plotone{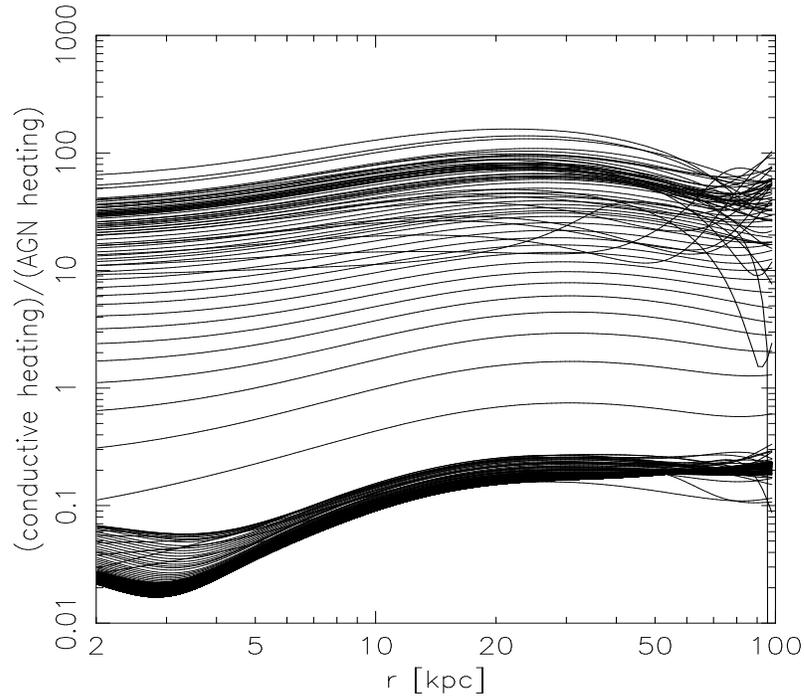}
\caption{Time sequence showing ratio of conductive heating rate to local AGN 
heating rate, as a function of radius, for the model displayed in 
Fig.~\ref{Fig2}. Conduction coefficient is set to 0.23 of Spitzer value. 
Although conduction dominates to begin with, it offsets only $\sim 10-15 \%$ 
of the radiative cooling in the final state; the rest is due to the AGN. }
\label{Fig3}
\end{figure}

The most important property of the above ``effervescent heating" rate is 
its proportionality to the pressure gradient (among other factors), since this 
determines the rate at which $pdV$ work is done as the bubbles rise.  Since 
thermal gas that suffers excess cooling will develop a slightly higher 
pressure gradient, the effervescent heating mechanism targets exactly those 
regions where cooling is strongest.  Therefore, it has the potential 
to stabilize radiative cooling (Begelman 2001b). This potential is borne 
out in 1D, time-dependent numerical simulations (Ruszkowski \& Begelman 2002), 
which show that the flow settles down to a steady state that resembles 
observed clusters, for a wide range of parameters and without fine-tuning the 
initial or boundary conditions (Fig.~\ref{Fig2}). Even though these models 
include conduction (at 23\% of the Spitzer rate), which may be necessary for 
global stability (Zakamska \& Narayan 2003), the heating is overwhelmingly 
dominated  by the AGN ($\geq 4:1$, at all radii; see Fig.~\ref{Fig3}). The 
mass inflow rate through the inner boundary, which determines the AGN feedback 
in these simulations, stabilizes to a reasonable value far below that 
predicted by cooling flow models. 

\subsection{The ``Entropy Floor"}

AGN feedback may do more than offset radiative losses in the cores of certain 
clusters. Cluster X-ray luminosities and gas masses increase with temperature 
more steeply than predicted by hierarchical merging models (Markevitch 1998; 
Nevalainen, Markevitch, \& Forman 2000).  In other words, the atmospheres in 
less massive clusters and groups are hotter than they should be, given 
the gravitational interactions that assembled them. These correlations apply 
to regions of clusters well outside the cooling radius, as well as to clusters 
without cooling cores.  They can be interpreted as evidence for an entropy 
``floor" (Lloyd-Davies, Ponman, \& Cannon 2000), indicating that low-entropy 
material is removed either by cooling and mass dropout (presumably to form 
stars: Bryan 2000; Voit et al. 2002) or by substantial AGN heating before or 
during cluster assembly (e.g., Valageas \& Silk 1999; Nath \& Roychowdury 
2002; McCarthy et al. 2002, and references therein).  

Mechanisms like those described above should also operate on these larger 
scales. For example, if the pressure gradient inside the cooling radius is 
not too steep, the effervescent heating model predicts that a substantial 
fraction of the injected energy will escape to radii where the cooling time 
is longer than the Hubble time.  The calculations have not been done yet. 

\section{Feedback and the Growth of Supermassive Black Holes}

The correlations measured between black hole masses and the velocity 
dispersions and/or masses of their host galaxies' bulges (Magorrian et al. 
1998; Ferrarese \& Merritt 2000; Gebhardt et al. 2000) suggest a direct 
relationship between supermassive black hole (SMBH) formation and galaxy 
formation.  Inventories of quasar light (So\l tan 1982; Yu \& Tremaine 2002, 
and references therein) and the hard X-ray background (Fabian \& Iwasawa 1999) 
suggest that much of the black hole growth occurred through (radiatively 
efficient) accretion, rather than through hierarchical mergers of smaller 
black holes.  If even a few percent of the liberated energy emerged in kinetic 
form, as seems very likely, then there would have been more than enough energy 
to unbind the gas in the protogalactic host.  Thus, several authors have 
suggested that SMBHs limited their own growth, or even the growth of the host 
galaxy, by depositing this energy in their surroundings (e.g., Silk \& Rees 
1998; Blandford 1999; Fabian 1999). If the energy is deposited adiabatically, 
the feedback luminosity $L_f$ can unbind the gas provided that 
\begin{equation}
L_f > {\sigma^5 \over G} = 5 \times 10^{43} \sigma^5_{200} \ 
{\rm erg} \ {\rm s}^{-1} .
\end{equation}
This implies a limiting black hole mass
\begin{equation}
\label{silkrees}
M_\bullet \sim 10^8 \left( {L_f \over 0.004 L_E}\right)^{-1} \sigma^5_{200} 
\ M_\odot , 
\end{equation}
where $L_E$ is the Eddington limit.  SMBHs must have accreted a significant 
fraction of their masses at close to the Eddington limit (Blandford 1999; 
Fabian 1999), since statistical arguments suggest that black hole growth took 
only a few Eddington $e$-folding times ($\sim {\rm few} \times 40$ Myr), and 
since at least some supermassive black holes existed only 1--2 Gyr after the 
Big Bang. Arguments based on the quasar luminosity function suggest that the 
most massive black holes might even have grown at several times the Eddington 
rate (Begelman 2001a, 2002) or with a higher radiative efficiency than is 
normally assumed (Yu \& Tremaine 2001). 
 
The above estimate assumes weak radiative losses in the protogalactic medium. 
However, the protogalactic environment is likely to be even more 
cooling-dominated than cluster cores (White \& Rees 1978; Fabian 1999).  If 
cooling is important it would require more energy to unbind the gas, by a 
factor that could be as large as $c / \sigma = 1500 \sigma_{200}^{-1}$.  The 
limiting mass is then 
\begin{equation}
\label{silkrees}
M_\bullet \sim 6 \times 10^8 \left( {L_f \over L_E}\right)^{-1} \sigma^4_{200} 
\ M_\odot . 
\end{equation}
The difference between these two limiting cases is analogous to the difference 
between the expansion of a supernova remnant in the energy-conserving Sedov 
(blast wave) phase, and its rapid deceleration during the momentum-conserving 
(radiative) ``snowplow" phase.  Reality is likely to be somewhere in between, 
with the flow of AGN energy partially trapped by density inhomogeneities that 
result from rapid cooling.

\section{Conclusions}

AGN feedback effects, potentially enormous on the basis of energetic 
arguments, depend sensitively on both the form of feedback and the detailed 
structure of the environment.  The efficiency of feedback due to radiation is 
often small, except under particular circumstances.  Kinetic energy injected 
by the AGN tends to be trapped by the ambient medium, leading to a higher 
efficiency. Recent theoretical advances suggest that accreting black holes 
often return a large fraction of the liberated energy to the environment in 
the form of winds and jets. 

The X-ray emitting atmospheres in clusters of galaxies provides an excellent 
testbed for the effects of AGN feedback.  Recent X-ray observations show that 
radio galaxies can blow long-lasting ``holes" in the ICM, and may offset the 
effects of radiative losses well enough to hamper large-scale inflows. Further 
observations, and sophisticated numerical simulations, will be needed to 
fully understand these interactions.

\bigskip

This work was supported in part by NSF grant AST--9876887.  I thank Mateusz 
Ruszkowski for many useful discussions, and for supplying Fig.~\ref{Fig3}.

\end{document}